%% file: report-new.tex
\documentclass{llncs}
\usepackage{amsmath}
\usepackage{stmaryrd}
\usepackage{listings}
\usepackage{times}
\usepackage{latexsym,amsmath,epsfig,amssymb,graphics,tabularx}
\usepackage{color,graphicx}

\usepackage{pgf}
\usepackage{tikz}
\usetikzlibrary{arrows,automata}
\usetikzlibrary{shapes}
\usetikzlibrary{positioning}
\tikzset{
    pshape/.style={
           rectangle,
           rounded corners,
           draw=black, very thick,
           minimum height=2em,
           inner sep=2pt,
           text centered,
           },
}

\tikzset{
    dshape/.style={
           rectangle,
           draw=black, very thick,
           minimum height=2em,
           inner sep=2pt,
           text centered,
           },
}

\lstdefinelanguage{isa} {alsoletter={\#, \&, $},
  mathescape=true,
  basicstyle=\small,
  escapechar=\@,
  boxpos=c,
  morekeywords= {and, axioms, axiomatization, Skip, Stop, type_synonym, theorem, lemma, apply, by, constdefs, definition, where,
    infixl, types, consts, primrec, have, from, show, let,  proof, qed, is,
    sorry, with, assume, fix, thus, hence, datatype, if, then, else, in, case,
    of
  },
  emph={[2] real, int, char, bool, string, self, boolean, Ass, Send, Rec, Seq, Cond, Pref,
                Join, Meet, Par, Rep, Cont, TOut, Inrp},
  emphstyle={[2]\it},
literate=
  {->}{{$\rightarrow$}}2
  {=>}{{$\Rightarrow$}}2
  {-->}{{$\Rightarrow$}}2
  {\\forall}{{$\forall$}}2
  {\\exist}{{$\exists$}}2
  {\\exists}{{$\exists$}}2
  {AND}{{$\& \&$}}3
  {ALL}{{$\forall$}}2
  {EX}{{$\exists$}}2
  {\%}{{$\lambda$}}1
  {\\\/}{{$\sqcap$}}2
  {|-}{{$\vdash$}}2
  {==}{{$\equiv$}}2
  {==>}{{$\Rightarrow$}}2
  {~}{{$\neg$}}1
  {~=}{{$\neq$}}2
  {:=}{{\tt :=}}2
  {;}{{\tt ;}}1
  {$}{{\do}}1
}

\lstnewenvironment{isaenv} {
  \lstset{extendedchars=true,columns=flexible}
  \lstset{language=isa,deletestring=[b]', basicstyle=\footnotesize}%
} {}

\newcommand{\pskip}{\textmd{skip}}
\newcommand{\pwait}{\textmd{wait}}
\newcommand{\evolution}[3]{\langle \mathcal{#1}(\dot{#2},#2)=0 \& #3\rangle}
\newcommand{\bexempt}[3]{#1 \unrhd #2 \rightarrow #3}

\newcommand{\leadm}[1]{\xrightarrow{#1}}

\newcommand{\pstop}{\epsilon}
\newcommand{\now}{now}
\newcommand{\Define}{\stackrel{\mbox{\small\rm def}}{=}}
\newcommand{\fracN}[2]{\frac{\small \textstyle #1}{\small \textstyle #2}}
\newcommand{\seman}[1]{[\![#1 ]\!]}
\newcommand{\la}{\ensuremath{\langle}}
\newcommand{\ra}{\ensuremath{\rangle}}
\newcommand{\ptrue}{true}
\newcommand{\pfalse}{false}
\newcommand{\ltrue}{\textmd{True}}
\newcommand{\chan}{\textbf{Chan}}
\newcommand{\dom}{\textrm{dom}}

\newcommand{\chop}{\smallfrown}

\newcommand{\typeb}[2]{\vdash #1 \blacktriangleright #2}

\newcommand{\soundb}[2]{\models #1 \blacktriangleright #2}

\newcommand{\semant}[1]{[\!\{#1\}\!]}
\newcommand{\rec}[1]{\langle\!|(#1)|\!\rangle}

\newcommand{\inference}[3]{\{#1\} \ #2\ \{#3\}}
\newcommand{\dceil}[1]{\lceil #1 \rceil}

\newcommand{\close}[1]{cl(#1)}

\newcommand{\Pre}{\varphi}
\newcommand{\Post}{\psi}
\newcommand{\HF}{HF}

\newcommand{\oomit}[1]{}

\begin{document}

\title{A Framework for Hybrid Systems with Denial-of-Service Security Attack}
\author{Shuling Wang\inst{1} \and Flemming Nielson\inst{2}
\and Hanne Riis Nielson \inst{2}
  }
   \institute{
 State Key Laboratory of Computer Science,
 Institute of Software \\
 Chinese Academy of
 Sciences, China
 \and DTU Informatics,
 Technical University of Denmark, Denmark}

\maketitle

\begin{abstract}
Hybrid systems are integrations of discrete computation and continuous physical evolution. The physical components of such systems
introduce safety requirements, the achievement of which asks for the correct monitoring and
control from the discrete controllers. However, due to denial-of-service security attack,
the expected information from the controllers is not received and
as a consequence the physical systems may fail to behave as expected. This paper
proposes a formal framework for expressing denial-of-service  security attack
 in hybrid systems.  As a virtue, a physical system is able to
plan for reasonable behavior in case the ideal control fails due to unreliable communication,
 in such a way that the safety of the system upon denial-of-service is still guaranteed.
 In the context of the modeling language, we develop an inference system
for verifying safety of hybrid systems, without putting any assumptions on how the environments behave.
Based on the inference system, we implement an interactive theorem prover and have applied
it to check an example taken from train control system.
\end{abstract}
\section{Introduction}

Hybrid systems, also known as cyber-physical systems, are dynamic systems
with interacting continuous-time physical systems and discrete controllers.
The physical systems evolve continuously with respect to time, such as
aircrafts, or  biological cell growth,  while
the computer controllers, such as autopilots, or biological control circuits, monitor and control
the behavior of the systems to meet the given design requirements. One design requirement is safety, which
includes time-critical constraints,  or invariants etc.,  for the example of train control systems,
the train should arrive at the stops on time, or
the train must always move with a velocity within a safe range.

However, due to the uncertainty in the environment, the potential errors in wireless communications between
the interacting components will make the safety of the system very hard to guarantee.
For the sake of safety, when the controllers fail to behave as expected because of absence of expected communication and thus become unreliable,
the physical systems should provide feedback control,
to achieve safety despite errors in communication. 


\paragraph{\textbf{A Motivating Example}}

\begin{figure}[htbp]
\centering
\small
\input{pic1}
\caption{The structure of train control example}
\label{fig:idea}
\end{figure}
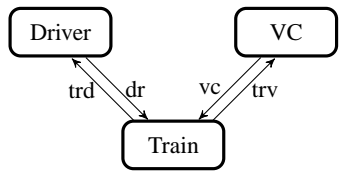

We illustrate our motivation by an example taken from train control system,
that is depicted in Fig.~\ref{fig:idea}.
It consists of three inter-communicating components:
Train, Driver and on board vital computer (VC).
We assume that the train owns arbitrarily long
movement authority, within which the train is allowed to move only,
 and must conform to a safety requirement, i.e.
the velocity must be non-negative and cannot exceed a maximum limit.
The train acts as a continuous plant, and moves with a given acceleration;
both the driver and the VC act as controllers, in such a way that,
either of them observes the velocity of the train periodically, and then according to the safety requirement,
computes the new acceleration for the train to follow in the next period.
According to the specification of the system, the message from the VC always takes high priority over the one from the driver.
\vspace{-2em}


\vspace{-1.5em}
However, the expected monitoring and control from VC or driver may fail due to denial-of-service security attack, e.g.
if the driver falls asleep, or if the VC
gets malfunction, and as a consequence, the  train may get no response from any of them within a duration of time.
The safety requirement of the train will then be violated very easily.
This  poses the problem of how to build a safe hybrid system in the presence of this sort of
denial-of-service security attack  from the environment.

The contribution of this paper includes the following aspects:
\begin{itemize}
  \item a programming notation, for formally modeling hybrid systems and meanwhile being able to express
          denial-of-service due to unreliable communications, and an assertion language, for describing safety as
          annotations in such programs;
  \item a deductive inference system, for reasoning about whether the program satisfies the annotated safety property, and a subsequent
           interactive theorem prover.
\end{itemize}
As a direct application, we are able to build a safe system for the example such that:
\begin{itemize}
\item[(F1)] the error configurations where neither driver nor VC is available are not reachable;
\item[(F2)] the velocity of the train keeps always in the safe range, although in the presence of denial-of-service
attack from the driver or the VC.
\end{itemize}
Furthermore, when the behavior of the environments (i.e. driver and VC) is determined,
e.g. by defining some constraints among the constants of the whole system, we can learn
 more precise behavior of the train.


In Section~\ref{sec:syntax} and Section~\ref{sec:semantics}, we
present the syntax and semantics for the formal modeling language. It is a combination of
Hybrid CSP (HCSP)~\cite{Jifeng:1994,Zhou:1996}, a process algebra based modeling language for describing hybrid systems,  and the binders
from Quality Calculus~\cite{RNV12}, a process calculus that allows one to take measures in case of unreliable communications.
With the introducing of binders, the modelling language is capable of programming
a safe system that executes in an open environment
that does not always live up to expectations.

In Section~\ref{sec:inference}, we define an inference system for reasoning about HCSP with binders.
For each construct $P$, the specification is of the form $\inference{\Pre}{P}{\Post, \HF}$,
where $\Pre$ and $\Post$ are the pre-/post-condition recording the initial and terminating states of
$P$ respectively,  and $HF$ the history formula recording the whole
execution history of $P$ (thus able to specify global invariants).
As a direct application, the (un-)reachability analysis
can be performed
by identifying the points corresponding
to the error configurations by logical formulas and then checking the (un-)satisfiability of the formulas.
In  Section~\ref{sec:application}, we have applied a theorem prover we have implemented to verify properties (F1) and (F2) of the train control example.
At last, we conclude the paper and address some future work.

\paragraph{\textbf{Related Work} }

There have been numerous work on formal modeling and verification of hybrid systems, e.g., \cite{Alur:1992,Manna93,Henzinger96,LSVW96,LPY02}, the most popular of which is {\em hybrid automata}~\cite{Alur:1992,Manna93,Henzinger96}. For
 automata-based approaches, the verification of hybrid systems is reduced to computing reachable sets,
 which is conducted either by model-checking~\cite{Alur:1992} or by the decision procedure of Tarski algebra~\cite{LPY02}. However,
  hybrid automata, analogous to state machines, has
  little support for structured description; and meanwhile,
   the verification of it based on reachability computation is not scalable and
   only applicable to some specific linear hybrid systems,  as
   it heavily depends on the decidability of the problems to be solved.
Applying abstraction or (numeric) approximation \cite{EClarke03,ADI06,ABDM00}  can improve the scalability,
but as a pay we have to sacrifice the precision.

 In contrast, deductive methods increasingly attract more attention in the verification
   of hybrid systems as it can scale up to complex systems.
  A differential-algebraic dynamic logic for hybrid programs \cite{Platzer10} was proposed by extending
dynamic logic with continuous statements, and has been applied for safety checking of European Train Control
System~\cite{PQ09}. However, the hybrid programs there can be considered as a textual encoding of hybrid automata,
with no support for communication and concurrency.
In~\cite{LLQZ10,ZWZictac}, the Hoare logic is extended to hybrid systems
modeled by Hybrid CSP~\cite{Jifeng:1994,Zhou:1996}, and then used for safety checking of
Chinese Train Control System. But the logic lacks compositionality.

All the work mentioned above focus on safety without considering denial-of-service security attacks from
the environment.
Quality Calculus~\cite{RNV12,HNN13} for the first time proposed a programming notation for expressing
denial-of-service in communication systems, but is currently limited to discrete time world.

\section{Syntax}
\label{sec:syntax}
 We first choose Hybrid CSP (HCSP)~\cite{Jifeng:1994,Zhou:1996} as the modelling language
 for hybrid systems.
 HCSP inherits from CSP the explicit communication model and concurrency, thus is expressive enough for describing
  distributed components and the interactions between them. Moreover, it extends CSP with differential
equations for representing continuous evolution, and provides several forms of interrupts to continuous evolution for
realizing communication-based discrete control.
On the other hand, Quality Calculus~\cite{RNV12,HNN13} is recently proposed to programming software components and their interactions in the presence
of unreliable communications. With the help of  \emph{binders} specifying the success or failure of communications and then the
communications to be performed before continuing,
it becomes natural in Quality Calculus to plan for reasonable behavior  in case the ideal
behavior fails due to unreliable communication and thereby to increase the quality of  the system.

In our approach, we will extend HCSP further with the notion of binders from Quality Calculus,
for modelling hybrid systems in the presence of denial-of-service because of unreliable communications.
The overall modelling language is given by the following syntax:
\vspace{-0.5em}
 \[
\begin{array}{lll}
e &::=& c  \mid x \mid f^k(e_1, ..., e_k)\\
 b &::=& ch!e\{u_1\} \mid ch?x\{u_2\} \mid \&_q(b_1, \cdots, b_n)\\
 P, Q  & ::= & \pskip \mid  x :=e  \mid b \mid  \evolution{F}{s}{B} \mid \bexempt{\evolution{F}{s}{B}}{b}{Q} \mid \\
      &&P\|Q \mid P;Q  \mid \omega \rightarrow P\mid  P^*
 \end{array}
 \]

Expressions $e$ are used to construct data elements and consist of constants $c$,
data variables $x$, and function application $f^k(e_1, ..., e_k)$.

Binders $b$ specify the inputs and outputs to be performed before continuing.
The output  $ch!e\{u_1\}$ expects to send message $e$ along channel $ch$, with
$u_1$ being the acknowledgement in case the communication succeeds, and the
dual input $ch?x\{u_2\}$ expects to receive a message from $ch$ and assigns it to variable $x$, with
$u_2$  being the acknowledgement similarly. We call both $u_1$ and $u_2$ \emph{acknowledgment variables}, and
assume in syntax that for each input or output statement, there exists a unique acknowledgement variable attached to it.
In the sequel, we will
use $\mathcal{V}$ and $\mathcal{A}$ to represent the set of data variables and acknowledgement variables respectively,
 and they are disjoint.
For the general form $\&_q(b_1, \cdots, b_n)$, the quality predicate $q$ specifies the sufficient communications
among
 $b_1, \cdots, b_n$  for the following process to proceed. In syntax, $q$ is a logical
combination of quality predicates corresponding to $b_1$, $\cdots$, $b_n$ recursively (denoted by $q_1, \cdots, q_n$ respectively below).
For example, the quality predicates for $ch!e\{u_1\}$ and
$ch?x\{u_2\}$ are boolean formulas $u_1  = 1$ and $u_2 = 1$.
There are two special forms of quality predicates, abbreviated as $\exists$ and $\forall$, with the definitions:
$
  \forall \Define q_1 \wedge \cdots \wedge q_n$ and $\exists \Define q_1 \vee \cdots \vee q_n
$.
More forms of quality predicates can be found in~\cite{RNV12}.

\example For the train example, define binder $b_0$ as
$\&_\exists(\textsf{dr}?x_a\{u_a\}, \textsf{vc}?y_a\{w_a\})$,
the quality predicate
of which amounts to $u_a=1 \vee w_a = 1$.
It expresses that,
the train is waiting for the acceleration from the driver and the VC, via \textsf{dr} and \textsf{vc} respectively,
and
as soon as one of the communications succeeds (i.e., when the quality predicate becomes true),
the following process will be continued without waiting for the other. $\qed$

 $P, Q$ define processes. The
$\pskip$ and assignment $x: = e$ are defined as usual, taking no time to complete.
Binders $b$ are explained above.
The continuous evolution $\evolution{F}{s}{B}$, where $s$ represents a vector of continuous variables and
$\dot{s}$ the corresponding  first-order derivative of $s$,
forces  $s$ to evolve according to the differential equations $\mathcal F$ as long as $B$,
a boolean formula of $s$ that  defines the {\em domain of $s$}, holds, and terminates when $B$ turns false. The communication interrupt $\bexempt{\evolution{F}{s}{B}}{b}{Q}$ behaves as $\evolution{F}{s}{B}$ first,
and if $b$ occurs before the continuous terminates,  the continuous  will be preempted  and
$Q$ will be executed instead.

The rest of the  constructs define compound processes. The parallel composition $P\|Q$
behaves as if $P$ and $Q$ run independently
except that the communications along the common channels connecting $P$ and $Q$ are to be synchronized.
In syntax, $P$ and $Q$ in parallel are restricted not to
share variables, nor input or output channels.
The sequential composition $P; Q$ behaves as $P$ first, and if it terminates, as $Q$ afterwards.
The conditional $\omega \rightarrow P$ behaves as $P$ if $\omega$ is true, otherwise terminates immediately. The condition $\omega$ can be used for 
checking the status of data variables or acknowledgement, thus in syntax,
it is a boolean formula on data and acknowledgement variables (while for
the above continuous evolution,  $B$ is a boolean formula on only data variables). The
repetition $P^*$ executes $P$ for arbitrarily finite number of times.

It should be noticed that, with the addition of binders, it is able to derive a number of other
known constructs of process calculi, e.g., internal
and external choice~\cite{RNV12}.

\example
Following Example 1,  the following model
\[
\begin{array}{l}
t:=0; {}^1 \langle \dot{s} = v, \dot{v} = a, \dot{t} = 1 \& t<T \rangle \unrhd b_0 {}^2\rightarrow \\
\quad ( w_a=1{}^3 \rightarrow a:=y_a; w_a = 0 \wedge u_a = 1{}^4 \rightarrow a:=x_a;
 \ \ w_a = 0 \wedge u_a = 0{}^5 \rightarrow \pskip)
\end{array}
\]
denoted by $P_0$, expresses that, the train moves with velocity $v$ and acceleration $a$, and as soon as $b_0$ occurs within $T$ time units, i.e.
 the train succeeds to receive a new acceleration from either the driver or the VC,
 then its acceleration $a$ will be updated by case analysis. It can be seen that the acceleration from VC will be used in priority.
For later reference we have annotated the program with labels (e.g. 1, 2, etc.). $\qed$



  \section{Transition Semantics}
\label{sec:semantics}
We first introduce a variable $now$ to record the global time during process execution, and then
define the set $\mathcal{V}^+= \mathcal{V} \cup \mathcal{A} \cup \{now\}$.
A state, ranging over $\sigma, \sigma'$,  assigns a value to each variable in
$\mathcal{V} ^+$, and we will use $\Sigma$ to represent the set of states.
A flow, ranging over $h, h'$, defined on a closed time interval $[r_1, r_2]$ with $0 \leq r_1 \leq r_2$,
or an infinite interval $[r, \infty)$ with some $r \geq 0$,
assigns a state in $\Sigma$ to each point in the interval.
Given a state $\sigma$, an expression $e$ is evaluated to a value under $\sigma$, denoted by $\sigma(e)$ below.

Each transition relation has the form $(P, \sigma) \leadm{\alpha} (P', \sigma', h)$,
where $P$ is a process, $\sigma, \sigma'$ are states, $h$ is a flow, and $\alpha$ is an event. It represents that
starting from initial state $\sigma$, $P$ evolves into $P'$ and ends with state $\sigma'$ and flow $h$, while
performing event $\alpha$. When the above transition takes no time, it produces a point flow,
i.e. $\sigma(now) = \sigma'(now)$ and
$h = \{\sigma(now) \mapsto \sigma'\}$, and we will call the transition \emph{discrete} and write
$(P, \sigma) \leadm{\alpha} (P', \sigma')$
instead without losing any information.
The label $\alpha$ represents
events, which can be a discrete internal event,
like skip, assignment, evaluation of boolean conditions, or termination of a continuous evolution etc.,
uniformly denoted by $\tau$, or an external communication, like output $ch!c\{1\}$ or input $ch?c\{1\}$,
or an internal communication $ch\dag c\{1\}$, or a time delay $d$ for some positive $d $.
We call the events but the time delay \emph{discrete events}, and will use $\beta$ to range over them.

The transition relations for binders are defined in Table~\ref{semantic:binder}.
The input $ch?x\{u\}$ may perform an external communication
$ch?c\{1\}$,  and as a result  $x$ will be
bound to $c$ and $u$ set to $1$, or
it may keep waiting for  $d $ time. For
the second case, a flow $h_d$ over $[\sigma(now), \sigma(now)+d]$ is produced, satisfying that
 for any $t$ in the domain, $h_d(t) = \sigma[now \mapsto t]$, i.e. no variable but the clock $now$ in $\mathcal{V}^+$
is changed during the waiting period. Similarly, there are two rules for output $ch!e\{u\}$. Here
 $\sigma[\now +d] $ is an abbreviation for $\sigma[\now \mapsto \sigma(\now)+d]$.

Before defining the semantics of general binders,
we introduce two auxiliary functions. Assume
$(b_1, \cdots,  b_n)$ is
an intermediate tuple of binders that occurs during execution (thus some of $b_i$s might contain
$\pstop$), $q$ a quality predicate, and $\sigma$ a state. The function
$\seman{q}(b_1, \cdots,  b_n)$ defines the truth value of $q$ under $(b_1, \cdots,  b_n)$, which is
calculated by replacing each sub-predicate $q_i$ corresponding to $b_i$ in $q$ by $b_i \equiv \pstop$ respectively;
and  function $\rec{b_1, b_2, \cdots, b_n} \sigma$ returns a state
that fully reflects the failure or success of binders $b_1, \cdots, b_n$, and can be
constructed from $\sigma$ by setting
the acknowledgement variables
corresponding to the failing inputs or outputs among $b_1, \cdots, b_n$ to be $0$.
Based on these definitions, binder
$\&_q(b_1, \cdots, b_n)$ may keep waiting for $d$ time, if $q$ is false under
$(b_1, \cdots,  b_n)$, or perform a discrete event $\beta$ that is enabled for some $b_i$, or perform a $\tau$ transition and
 terminate if $q$ is true under $(b_1, \cdots,  b_n)$. Notice that when $q$ becomes true, the enabled discrete events  can still be performed, as indicated by the second rule.
\begin{table}[t]
\small
\centering
\begin{tabular}{c}
\hline \\
$(ch?x\{u\}, \sigma) \leadm{ch?c\{1\}} (\pstop, \sigma[x \mapsto c, u \mapsto 1])$ \\[0.5em]
$(ch?x\{u\}, \sigma) \leadm{d} (ch?x\{u\},\sigma[\now +d], h_d)$\\[0.5em]
$(ch!e\{u\}, \sigma) \leadm{ch!\sigma(e)\{1\}} (\pstop, \sigma[u \mapsto 1])$ $(ch!e\{u\}, \sigma) \leadm{d} (ch!e\{u\},\sigma[\now +d], h_d)$\\[0.5em]
\hline \\
$\seman{q}(b_1, \cdots,  b_n) = q[(b_1 \equiv \pstop)/q_1, \cdots,  (b_n \equiv \pstop)/q_n]$ \\[0.5em]
$\rec{} \sigma = \sigma \quad  \rec{\pstop, b_2, \cdots, b_n} \sigma =  \rec{b_2, \cdots, b_n} \sigma$\\[0.5em]
$\rec{ch?x\{u\}, b_2, \cdots, b_n} \sigma =  \rec{b_2, \cdots, b_n} (\sigma[u \mapsto 0]) $\\[0.5em]
$\rec{ch!e\{u\}, b_2, \cdots, b_n} \sigma =  \rec{b_2, \cdots, b_n} (\sigma[u \mapsto 0])$\\[0.5em]
$\rec{\&_{q_k}(b_{k1}, \cdots, b_{km}), b_2, \cdots, b_n} \sigma =\rec{b_{k1}, \cdots, b_{km}, b_2, \cdots, b_n}  \sigma $\\[0.5em]
\hline\\
$\fracN{\seman{q}(b_1, \cdots,  b_n) = \pfalse}{(\&_q(b_1, \cdots, b_n), \sigma) \leadm{d}
(\&_q(b_1, \cdots, b_n),\sigma[\now +d], h_d)}$\\[1em]
$\fracN{(b_i, \sigma) \leadm{\beta} (b_i', \sigma') }
{(\&_q(b_1, \cdots,  b_i, \cdots, b_n), \sigma) \leadm{\beta} (\&_q(b_1, \cdots, b_i', \cdots, b_n), \sigma')}$\\[1.2em]
$\fracN{
\seman{q}(b_1, \cdots, b_n) = \ptrue \quad \rec{b_1, \cdots,  b_n} \sigma = \sigma'}
{(\&_q(b_1, \cdots, b_n), \sigma) \leadm{\tau} (\pstop, \sigma')}$\\[1.3em]
\hline\\[-0.52em]
\end{tabular}
\caption{The transition relations for binders and the auxiliary functions}
\label{semantic:binder}
\end{table}

\example
Starting from  $\sigma_0$, the execution of $b_0$ in Example 1
may lead to three possible states at termination:
\begin{itemize}
\item
$\sigma_0[\now + d, x_a \mapsto c_a, u_a \mapsto 1, w_a \mapsto 0]$, indicating that the train succeeds to receive
$c_a$ from the driver after $d$ time units have passed, but fails for the VC;
\item $\sigma_0[\now + d, y_a \mapsto d_a, w_a \mapsto 1, u_a \mapsto 0]$, for the opposite case of the first;
\item $\sigma_0[\now + d, x_a \mapsto c_a, u_a \mapsto 1, y_a \mapsto d_a, w_a \mapsto 1]$, indicating that
the train succeeds to receive messages from the driver as well as the VC after $d$ time. $\ \qed$
\end{itemize}

\begin{table}[p]
\small
\centering
\begin{tabular}{c}
\hline \\
\multicolumn{1}{l}{\textbf{Skip, Assignment and Idle} $\quad$
${(\pskip, \sigma) \leadm{\tau} (\pstop, \sigma)}$}\\
${(x:=e, \sigma) \leadm{\tau} (\pstop, \sigma[x\mapsto \sigma(e)])} \quad$
$(\pstop, \sigma) \leadm{d} (\pstop, \sigma[\now +d], h_d)$
\\[1.3em]
\hline \\
\multicolumn{1}{l}{\textbf{Continuous Evolution} $\quad$ For any $d>0$, }\\
$\fracN{
\begin{array}{c}
 \mbox{ $S(t)$ is a solution of } \mathcal{F}(\dot{s},s)=0
  \mbox{ over $[0,d]$  satisfying that $S(0) = \sigma(s)$} \\
 \mbox{ and } \forall t\in [0, d).
 h_{d,s}(t + \sigma(now))(B) = \ptrue
\end{array}}
{
\begin{array}{c}
    (\la\mathcal{F}(\dot{s},s)=0 \& B \ra, \sigma) \leadm{d}
    (\la \mathcal{F}(\dot{s},s)=0 \& B \ra,
    \sigma[\now +d, s\mapsto S(d)], h_{d,s})
\end{array}}$
\\[2em]
$\fracN{
\begin{array}{c}
  (\sigma(B) =   \pfalse) \mbox{ or } (\sigma(B) =   \ptrue \wedge \exists \delta>0.\\
  (S(t) \mbox{ is a solution of } \mathcal{F}(\dot{s},s)=0 \mbox{  over }
  \mbox{ $[0, \delta]$ satisfying that $S(0) = \sigma(s)$ } \\
  \mbox{and } \forall t\in (0, \delta). h_{\delta,s}(t + \sigma(now))(B) = \pfalse))
\end{array}}
{(\la \mathcal{F}(\dot{s},s)=0 \& B \ra, \sigma) \leadm{\tau}
 (\pstop, \sigma )}$
\\[1.6em]
\hline\\
\multicolumn{1}{l}{\textbf{Communication Interrupt}}\\
$\fracN{
\begin{array}{c}
  (\evolution{F}{s}{B}, \sigma) \leadm{d} (\evolution{F}{s}{B},
  \sigma', h)
  \quad (b, \sigma) \leadm{d}  (b, \sigma'', h'')
  \end{array}}
  {
  \begin{array}{l}
  (\bexempt{\evolution{F}{s}{B}}{b}{Q}, \sigma)
   \leadm{d} (\bexempt{\evolution{F}{s}{B}}{b}{Q}, \sigma', h)
   \end{array}}$
\\[1.3em]
$\fracN{(b, \sigma) \leadm{\beta} (b', \sigma') \quad b' \neq \pstop}
  {
  \begin{array}{l}
(\bexempt{\evolution{F}{s}{B}}{b}{Q}, \sigma) \leadm{\beta} (\bexempt{\evolution{F}{s}{B}}{b'}{Q},\sigma')
  \end{array}}$
\\[1.3em]
$\fracN{(b, \sigma) \leadm{\beta} (\pstop, \sigma') }
  {(\bexempt{\evolution{F}{s}{B}}{b}{Q}, \sigma) \leadm{\beta} (Q, \sigma')}$\\[1.3em]
$  \fracN{
\begin{array}{c}
  (\evolution{F}{s}{B}, \sigma) \leadm{\tau} (\pstop, \sigma')
   \quad \neg((b, \sigma) \leadm{\tau} (\pstop, -))\\
   b \equiv \&_q(b_1, \cdots, b_n) \quad \rec{b_1, \cdots,  b_n} \sigma' = \sigma''
\end{array}}
   { (\bexempt{\evolution{F}{s}{B}}{b}{Q}, \sigma) \leadm{\tau}(\pstop, \sigma'')}$
\\[1.6em]
\hline\\
\multicolumn{1}{l}{\textbf{Parallel Composition}}\\
$\fracN{(P, \sigma_1) \leadm{ch?c\{1\}} (P', \sigma_1')  \quad
     (Q, \sigma_2)\leadm{ch!c\{1\}} (Q', \sigma_2')}
  {(P \parallel Q, \sigma_1\uplus \sigma_2) \leadm{ch \dag c \{1\}}  (P'||Q', \sigma_1'\uplus \sigma_2')}$
\quad\\[1.3em]
$\fracN{
 \begin{array}{c}
 (P, \sigma_1) \leadm{\beta} (P', \sigma_1') \quad
 \beta \in \{\tau, ch \dag c \{1\},
 ch ? c \{1\}, ch ! c \{1\} \mid \\
 ch \notin \chan(P) \cap \chan(Q) \} \quad
 \forall ch, c .  (\neg ((P,\sigma_1) \leadm{ch?c\{1\}} \wedge
       (Q,\sigma_2) \leadm{ch!c\{1\}}) \\
       \wedge \neg ( (P,\sigma_1) \leadm{ch!c\{1\}} \wedge
       (Q,\sigma_2) \leadm{ch?c\{1\}}))
 \end{array}}
  {(P \parallel Q, \sigma_1\uplus \sigma_2) \leadm{\beta}  (P'||Q, \sigma_1'\uplus \sigma_2)}$
\\[1.3em]
$\fracN{
  \begin{array}{c}
     (P, \sigma_1) \leadm{d} (P', \sigma_1', h_1) \quad
     (Q, \sigma_2) \leadm{d} (Q', \sigma_2', h_2)   \\
     \forall ch, c .  \neg ((P \parallel Q, \sigma_1 \uplus \sigma_2) \leadm{ch \dag c \{1\}}) \quad
       \neg ((P \parallel Q, \sigma_1 \uplus \sigma_2) \leadm{\tau})
  \end{array}}
  {(P \parallel Q, \sigma_1 \uplus \sigma_2) \leadm{d}  (P'||Q', \sigma_1'\uplus \sigma_2', h_1 \uplus h_2)}$
  \\
  $(\pstop \| \pstop, \sigma) \leadm{\tau} (\pstop, \sigma)$
\\
\hline\\
\multicolumn{1}{l}{\textbf{Other Compound Constructs}}\\
$\fracN{\sigma(\omega) = \ptrue}{(\omega\rightarrow P, \sigma) \leadm{\tau}(P,\sigma)}   \quad
\fracN{\sigma(\omega) = \pfalse} {(\omega\rightarrow P, \sigma) \leadm{\tau}(\pstop,\sigma)}$
\\[1.3em]
$\fracN{(P,\sigma) \leadm{\alpha} (P', \sigma', h) \quad P' \neq \pstop} {(P; Q, \sigma) \leadm{\alpha} (P';Q, \sigma', h)}$
 $\quad$
$\fracN{(P,\sigma) \leadm{\alpha} (\pstop, \sigma', h)  }{(P; Q, \sigma) \leadm{\alpha} (Q, \sigma', h)}$\\[1.3em]
$\fracN{(P, \sigma) \leadm{\alpha} (P', \sigma', h) \quad  P' \neq \pstop}{(P^*, \sigma)\leadm{\alpha} (P'; P^*, \sigma', h)} \quad
\fracN{(P, \sigma) \leadm{\alpha} (\pstop, \sigma', h) }{(P^*, \sigma)\leadm{\alpha} (P^*, \sigma', h)} \quad$
$(P^*, \sigma) \leadm{\tau} (\pstop, \sigma)$\\[1.6em]
\hline
 \end{tabular}
\caption{The transition relations for processes}
\label{semantic:process}
\end{table}

The transition relations for other processes are defined in Table~\ref{semantic:process}. The rules
for skip and assignment can be defined as usual. The idle rule represents that
the process can stay at the terminating state $\pstop$ for arbitrary $d$
time units, with nothing changed but only the clock progress.
 For continuous evolution, for any $d>0$,
it evolves for $d$ time units according to $\mathcal{F}$
if $B$ evaluates to true within this period (the right end exclusive).
A flow  $h_{d, s}$  over $[\sigma(now), \sigma(now)+d]$ will then be produced, such that
 for any $o$ in the domain,
 $h_{d, s}(o)  = \sigma[now \mapsto o, s \mapsto S(o-\sigma(now))]$, where
 $S(t)$ is the solution as defined in the rule.
 Otherwise, the continuous evolution
terminates at a point if $B$ evaluates to false at the point, or if $B$ evaluates to false
 at a positive open interval right to the point.

For communication interrupt, the process may evolve for $d$ time units if both the continuous evolution and
the binder can progress for $d$ time units, and then reach the same state and flow as the continuous evolution does.
It may perform a discrete event over $b$, and if the resulting binder $b'$ is not
$\pstop$, then the continuous evolution is kept, otherwise,
the continuous evolution will be interrupted and $Q$ will be followed to execute, and for both cases,
will reach the same state and flow as the binder does.
Finally, it may perform a $\tau$ event and terminate immediately
 if the continuous evolution terminates with a $\tau$ event but $b$ not.
Notice that  the final state $\sigma''$ needs to be reconstructed from $\sigma'$
 by resetting the acknowledgement variables of those unsuccessful  binders occurring in $b$
 to be $0$.

Before defining the semantics of  parallel composition, we need to introduce some notations.
Two states $\sigma_1$
and $\sigma_2$ are \emph{disjoint}, iff $\dom(\sigma_1) \cap \dom(\sigma_2) = \{now\}$ and $\sigma_1(now) = \sigma_2(now)$. For two disjoint states $\sigma_1$
and $\sigma_2$,  $\sigma_1 \uplus \sigma_2$ is defined as a state over
$\dom(\sigma_1) \cup \dom(\sigma_2) $, satisfying that
$\sigma_1 \uplus \sigma_2(v)$ is $\sigma_1(v)$  if  $v \in \dom(\sigma_1)$, otherwise
$\sigma_2(v)$  if  $v \in \dom(\sigma_2)$.
We lift this definition to flows $h_1$ and $h_2$ satisfying
$\dom(h_1) = \dom(h_2)$, and define $h_1 \uplus h_2$ to be a flow such that
$h_1 \uplus h_2 (t)=h_1(t) \uplus h_2(t)$.
For $P\|Q$, assume $\sigma_1$ and $\sigma_2$   represent the initial states for
$P$ and $Q$ respectively and are disjoint.
The process  will perform a communication along a common channel of $P$ and $Q$, if
$P$ and $Q$ get ready to synchronize with each other along the channel.
Otherwise, it will
perform a discrete event, that can be $\tau$, an internal
communication of $P$, or an external communication along some non-common channel of $P$ and $Q$,
if $P$  can progress separately
on this event (and the symmetric rule for $Q$ is left out here). When
neither internal communication nor $\tau$ event is enabled for $P||Q$, it
may evolve for $d$ time units if both $P$ and $Q$ can evolve for $d$ time units. Finally,
 the process will perform a $\tau$ event and terminate as soon as
both the components terminate.

At last, the rules for conditional, sequential,  and repetition are defined as usual.

\example
Starting from state $\sigma_0$, the execution
of $P_0$ in Example 2 leads to the following cases (let $v_0$ denote $\sigma_0(v)$ below):
\begin{itemize}
\item $P_0$ terminates without the occurrence of $b_0$,
the final state is $\sigma_0[now+T, t +T, v+aT, s+v_0T+0.5 a T^2, u_a \mapsto 0, w_a \mapsto 0]$;
\item $b_0$ occurs after $d$ time units for some $d \leq T$, and as a result $P_0$ executes to location 2, with state $\sigma_0[now+d, t +d, v+ad, s+v_0d+0.5 a d^2, u_a,  w_a, x_a, y_a]$, where  $u_a$, $w_a$, $x_a$ and $y_a$ have 3 possible evaluations as defined in Example 3,  and then depending on the values of $u_a$ and $w_a$, executes to location 3 or 4  respectively, and finally terminates after a corresponding acceleration update. $\ \qed$
\end{itemize}

\paragraph{\textbf{Flow of a Process}}
Given two flows $h_1$ and $h_2$ defined on $[r_1, r_2]$ and $[r_2, r_3]$ (or $[r_2, \infty)$)
respectively, we define the concatenation  $h_1 ^\chop h_2 $ as the
flow defined on $[r_1, r_3]$ (or $[r_1, \infty)$) such that $h_1 ^\chop h_2 (t)$ is equal to $h_1(t)$ if $t\in [r_1, r_2)$, otherwise
$h_2(t)$. Given a process $P$ and an initial state $\sigma$, if we have the following sequence of transitions:
\[
\begin{array}{l}
   (P, \sigma) \leadm{\alpha_0} (P_1, \sigma_1, h_1) \quad
   (P_1, \sigma_1) \leadm{\alpha_1} (P_2, \sigma_2, h_2)\\
   \ldots \quad
   (P_{n-1}, \sigma_{n-1}) \leadm{\alpha_{n-1}} (P_n, \sigma_n, h_n)
\end{array}\]
then we define $h_1 ^\chop \ldots ^\chop h_n$ as the \emph{flow} from $P$ to $P_n$ with respect to the initial state $\sigma$,
and furthermore, write
$(P, \sigma) \leadm{\alpha_0 \cdots \alpha_{n-1}} (P_n, \sigma_n, h_1 ^\chop \ldots ^\chop h_n)$
to represent the whole transition sequence (and for simplicity, the label sequence can be omitted sometimes).
When $P_n$ is $\pstop$, we call $h_1 ^\chop \ldots ^\chop h_n$ a \emph{complete flow} of $P$ with respect to  $\sigma$.

\section{Inference System}
\label{sec:inference}

In this section, we define an inference system for reasoning about both discrete and continuous properties of HCSP with binders,
which are considered for an isolated time point and a time interval respectively.

\paragraph{\textbf{History Formulas}}
In order to describe the interval-related properties,
we  introduce history formulas,  that are defined by duration calculus (DC)~\cite{ZHR91,ZH04}.
DC is a first-order interval-based
real-time logic with one binary modality known as chop $^\chop$. History formulas $HF$ are defined
by the following subset of DC:
\[
\begin{array}{lll}
  HF ::=  \ell \circ T \mid \dceil{S} \mid  HF_1 {}^\chop HF_2 \mid \neg HF \mid HF_1 \vee HF_2
\end{array}\]
where $\ell$ is a temporal variable denoting the length
of the considered interval, $\circ \in \{<, =\}$ is a relation,  $T$ a non-negative real,
and $S$  a first-order state formula over process variables. For simplicity, we will write
$\dceil{S}^<$ as an abbreviation for $\dceil{S} \vee \ell=0$.

$HF$ can be interpreted over flows and intervals. We define the judgement $h, [a, b] \models HF$ to
represent that $HF$ holds under $h$ and $[a, b]$, then we have
\[
\begin{array}{ll}
  h, [a, b] \models \ell \circ T \mbox{ iff } (b-a) \circ T\qquad
  h, [a, b] \models \dceil{S} \mbox{ iff } \int_{a}^{b} h(t)(S) = b-a\\
   h, [a, b] \models HF_1 {}^\chop HF_2 \mbox{ iff } \exists c. a \leq c \leq b \wedge h, [a, c] \models HF_1 \wedge   h, [c, b] \models HF_2
\end{array}\]
 As defined above,
$\ell$ indicates the length of the considered interval;
 $\lceil S \rceil$ asserts that
  $S$ holds almost everywhere in the considered interval;
and $HF_1 ^\frown HF_2$ asserts that the interval can be divided into two sub-intervals such that
$HF_1$ holds for the first and $HF_2$ for the second. The first-order connectives $\neg$ and
$\vee$ can be explained as usual.

All  axioms and inference rules for DC presented in~\cite{ZH04} can be applied here,  such as
\[
\begin{array}{c}
  \ltrue \Leftrightarrow \ell \geq 0 \quad \dceil{S} {}^\chop \dceil{S} \Leftrightarrow \dceil{S} \quad HF ^\chop \ell = 0 \Leftrightarrow HF\\
  \dceil{S_1} \Rightarrow \dceil{S_2} \mbox{ if $S_1 \Rightarrow S_2$ is valid in FOL}\\
   \end{array}\]

\paragraph{\textbf{Specification}}
The specification for process $P$ takes form
$\inference{\Pre}{P}{\Post, \HF}$,
where  the pre-/post-condition $\Pre$ and $\Post$, defined by FOL,
specify properties of
variables that hold at the beginning  and termination of the execution of $P$ respectively,
 and the history formula $\HF$, specifies properties of  variables
 that hold throughout the execution interval of $P$. The specification of $P$ is
 defined with no dependence on
the behavior of its environment.
The specification is \emph{valid}, denoted by
$\models \inference{\Pre}{P}{\Post, \HF}$,
iff for any state $\sigma$, if $(P, \sigma) \leadm{} (\pstop, \sigma', h)$, then $\sigma \models \Pre$ implies
$\sigma' \models \Post$ and $h, [\sigma(now), \sigma'(now)] \models \HF$.

 \paragraph{\textbf{Acknowledgement of Binders }}
In order to define the inference rules for binders $b$, we first define
an auxiliary typing judgement $\typeb{b}{\varphi}$,
where  the first-order formula $\varphi$ describes the acknowledgement corresponding to successful passing of $b$,
and is defined without dependence on the precondition of $b$.
We say $b \blacktriangleright \varphi$ \emph{valid},
denoted by $\soundb{b}{\varphi}$,  iff
given any state $\sigma$, if $(b, \sigma) \leadm{ } (\pstop, \sigma', h)$, then
$\sigma'\models \varphi$ holds.

The typing judgement for binders is defined as follows:
\[
\begin{array}{c}
\typeb{ch?x\{u\}}{u=1} \quad \typeb{ch!e\{u\}}{u=1} \quad
\fracN{\typeb{b_1}{\varphi_1}, \ \cdots ,\ \typeb{b_n}{\varphi_n}}{\typeb{\&_q(b_1, \cdots, b_n)}{\semant{q}(\varphi_1, \cdots, \varphi_n)}}
\end{array}\]
As indicated above, for input $ch?x\{u\}$, the successful passing of it gives rise to formula $u=1$, and similarly for
output $ch!e\{u\}$; for binder $\&_q(b_1, \cdots, b_n)$, it gives rise to
formula $\semant{q}(\varphi_1, \cdots, \varphi_n)$, which encodes the effect of quality
predicate $q$ to the sub-formulas $\varphi_1, \ldots, \varphi_n$ corresponding to $b_1, \ldots, b_n$ respectively.

\example
For binder $b_0$ in Example 1, we have
$\typeb{b_0}{u_a = 1 \vee w_a = 1} $,
indicating that, if the location after $b_0$ is reachable, then at least one of the communications with the driver or the VC succeeds. $\qed$

\subsection{Inference Rules}

\begin{table}[t]
\small
\centering
\begin{tabular}{c}
\hline \\
$\inference{\Pre}{\pskip}{\Pre, \ell=0} \quad \inference{\Post[e/x]}{x :=e}{\Post, \ell=0}$ \\[0.6em]
$\inference{\Pre}{ch?x\{u\}}{(\exists x, u. \Pre) \wedge u=1, \dceil{\Pre}^<} \ $
$\inference{\Pre}{ch!e\{u\}}{(\exists u. \Pre) \wedge u=1, \dceil{\Pre}^<} $\\[0.6em]
$\fracN{\typeb{\&_q(b_1, \cdots, b_n)}{\alpha}}{\inference{\Pre}{\&_q(b_1, \cdots, b_n)}
{
\begin{array}{l}
(\exists mv(\&_q(b_1, \cdots, b_n)). \Pre) \wedge \alpha,
\dceil{\exists mv(\&_q(b_1, \cdots, b_n)). \Pre}^<
\end{array}}}$
\\[1.2em]
$\inference{\Pre}{\evolution{F}{s}{B}}{
\begin{array}{l}
(\exists s. \Pre) \wedge \close{\neg B} \wedge \close{Inv},
\dceil{(\exists s. \Pre) \wedge B \wedge Inv}^<
\end{array}}$
\\[0.8em]
$\fracN{
\begin{array}{c}
\typeb{\&_q(b_1, \cdots, b_n)}{\alpha} \quad
\inference{(\exists mv(b). (\exists s. \Pre) \wedge \close{Inv}) \wedge \alpha}{Q}{\Post_1, h_1}
\end{array}}
{\begin{array}{l}
 \{\Pre\} \bexempt{\evolution{F}{s}{B}}{b}{Q}
 \begin{array}{l}
  \{(\exists mv(b). (\exists s. \Pre) \wedge \close{\neg B} \wedge \close{Inv}) \vee \Post_1, \\
\dceil{\exists mv(b). (\exists s. \Pre) \wedge B \wedge Inv}^< {}^\chop  (\ell=0 \vee h_1)\}
\end{array}
\end{array}}$
\\[1.2em]
$\fracN{\inference{\Pre}{P}{\Post_1, h_1} \quad \inference{\Pre}{Q}{\Post_2, h_2}}
{\inference{\Pre}{P\|Q}{\Post_1 \wedge \Post_2, ((h_1 ^\chop \ltrue)\wedge h_2) \vee (h_1 \wedge (h_2 ^\chop \ltrue))}} $\\[1.3em]
$\fracN{\inference{\Pre}{P} {\Post_1, h_1}\quad \inference{\Post_1}{Q} {\Post_2, h_2}}{\inference{\Pre}{P; Q}{\Post_2, h_1 ^\chop h_2}}$
\quad
$\fracN{\inference{\Pre \wedge \omega}{P}{\Post_1, h_1}}{\inference{\Pre}{ \omega \rightarrow P}{(\Pre\wedge \neg \omega) \vee \Post_1,  \ell=0\vee h_1}}$ \\[1.3em]
$\fracN{\inference{\Pre}{P}{\Pre, Inv} \quad Inv ^\chop Inv \Rightarrow Inv}{\inference{\Pre}{P^*}{\Pre, Inv \vee \ell=0}}$\\[1.3em]
\hline\\
\end{tabular}
\caption{An inference system for processes}
\label{inferencesystem}

\end{table}

We first introduce an auxiliary  function
$mv(b)$, which given a binder $b$, returns the variables that may be modified by $b$. It can be defined directly  by structural induction on $b$ and we omit the details.
The inference rules for deducing the
 specifications of all constructs are presented in Table~\ref{inferencesystem}.

%

Statements $\pskip$ and assignment are defined as in classical Hoare Logic, plus
$\ell=0$ in the history formula, indicating that they both take zero time to complete.
For each form of the binders $b$, the postcondition is the conjunction of
the quantified precondition $\Pre$ over variables in
$mv(b)$ and  the acknowledgement corresponding to the successful
passing of $b$. The binders  may occur without waiting any time,
indicated by $\ell =0$ as one disjunctive clause of each history formula.
For both $ch?x\{u\}$ and $ch!e\{u\}$, if the waiting time is greater than
0, then  $\Pre$ will hold almost everywhere in the waiting interval
 (the only possible exception is the right endpoint, at which the communication occurs and variables might be changed correspondingly). For
 $\&_q(b_1, \cdots, b_n)$,  only the quantified $\Pre$  over  variables in
$mv(b)$ is guaranteed to hold almost everywhere throughout the waiting interval, since
some binders $b_i$s that make $q$ true might occur at sometime during the interval
and as a consequence variables in $\Pre$ might get changed.

For continuous evolution, the notion of differential invariants is used instead of explicit solutions.
A \emph{differential invariant} of $\evolution{F}{s}{B}$
for given initial values of $s$ is a first-order formula of $s$, which is satisfied by the initial values and also by all the values
reachable by the trajectory of $s$ defined by $\mathcal{F}$ within the domain $B$.
A method on generating differential invariants for polynomial differential equations was
proposed in~\cite{LZZ11}.
Here we assume $Inv$ is a differential invariant with respect to precondition $\Pre$ for the continuous evolution (more details on using
$Inv$ are shown in the later example proof). For the postcondition,
the quantified $\Pre$ over  the only
modified variables $s$, the closure of $\neg B$, and the closure of
$Inv$ hold. The closure $\close{\cdot}$  extends the domain defined by the corresponding formula to include the boundary.
For the history formula,  the execution interval may be 0, or otherwise, the quantified $\Pre$ over $s$, $B$ and $Inv$ holds almost everywhere throughout the
 interval.

For communication interrupt,
if $b$ fails to occur before the continuous evolution terminates,
the effect of the whole statement is almost equivalent to  the continuous  evolution,
except that some variables in $b$ may get changed because of occurrences of some communications
during the execution of the continuous  evolution.
Otherwise, if $b$ succeeds within the termination of the continuous  evolution,
the continuous evolution will be interrupted and $Q$ will start to execute from the interrupting point.
At the interrupting point, the acknowledgement of $b$ holds, and moreover,
because  $s$ and  variables in $mv(b)$ may have been modified,
$\exists mv(b). ((\exists s. \Pre) \wedge \close{Inv})$  holds (the closure here is to
include the case when the interrupting point is exactly the termination point of the continuous  evolution).
For the second case, the postcondition
is  defined as the one for $Q$, and the history formula as the chop of the one for
the continuous evolution before interruption and the one for $Q$ afterwards.
Finally, as indicated by the rule, the postcondition and history formula for the whole statement are defined as the disjunction of the above two cases.

The rule for $P\|Q$ is defined by conjunction, however, because
$P$ and $Q$ may terminate at different time,
the formula $\ltrue$ is added to the end of the history formula with short time interval to make
the two intervals equal. For $P;Q$, the history formula is defined by the concatenation of the ones
of $P$ and $Q$. The rule for $\omega \rightarrow P$ includes two cases depending on whether
$\omega$ holds or not.
At last, for $P^*$, we need to find the invariants, i.e. $\Pre$ and $Inv$,  for both the postcondition and history formula.

The general inference rules that are applicable to all constructs, like monotonicity, case analysis etc.,
can be defined as usual and are omitted here.

We have proved the following soundness theorem:
\begin{theorem}
Given a process $P$, if $\inference{\Pre}{P}{\Post, \HF}$ can be deduced from the inference rules,
then $\models \inference{\Pre}{P}{\Post, \HF}$.
\end{theorem}
 \textsc{proof}.
We need to prove that, for any state $\sigma$, if $(P, \sigma) \leadm{} (\pstop, \sigma', h)$, then $\sigma \models \Pre$ implies
$\sigma' \models \Post$ and $h, [\sigma(now), \sigma'(now)] \models \HF$. The proof is given by structural induction on $P$ as follows.
  \begin{itemize}
    \item The proof for  $\pskip$ and $ x:=e$ is trivial.

    \item Cases binders $b$: For $b \equiv ch?x\{u\}$,  according to the transition system, there exist some $d\geq 0$ and $c$ such
    that
    $\sigma' = \sigma[\sigma(now) \mapsto \sigma(now) + d][x\mapsto c, u \mapsto 1]$ and $h$
    defined on $[\sigma(now), \sigma(now) + d]$ satisfies that $h(t) = \sigma[now \mapsto t]$ for
    each $t$ in $[\sigma(now), \sigma(now) + d)$ and $h(\sigma(now) + d) = \sigma'$.
    Thus, from $\sigma \models \Pre$, $\sigma' \models \exists x, u. \Pre$
    and $h, [\sigma(now), \sigma'(now)]  \models \dceil{\Pre}^<$ must hold (notice that $now$ does not occur
    in assertions). The case for $b \equiv ch!e\{u\}$
    can be proved similarly.

  For $b \equiv \&_q(b_1, \cdots, b_n)$, according to the transition system,
    there must exist some $d\geq 0$ such that $\sigma'(now) = \sigma(now) + d$, and
    for each $b_i$ evolving to $\pstop$ at termination, there must be $\sigma'(u_i) = 1$, and
    for any variable $x$ that is not $mv(b)$, for any
    $t \in [\sigma(now), \sigma(now')]$, $h(t)(x) = \sigma(x)$. Thus
    $\sigma' \models \exists mv(b) . \Pre$ and $h, [\sigma(now), \sigma'(now)]  \models
    \dceil{\exists mv(b) . \Pre}^<$ hold. And, from $\seman{q}(b_1', \cdots, b_n') = \ptrue$, where $b_1', \cdots, b_n'$ represent the final form of $b_1, \cdots, b_n$ during the execution of $b$, we have $\sigma' \models \alpha$ proved.

    \item Case $\la\mathcal{F}(\dot{s},s)=0 \& B \ra$: According to the transition system,
    there must exist $d \geq 0$ such that $\sigma'=\sigma[\now \mapsto \sigma(\now)+d, s\mapsto S(d)]$  and
    $h$ defined over $[\sigma(now), $\\
    $\sigma(now)+d]$ satisfies that
    for any $o$ in the domain,  $h(o)  = \sigma[now \mapsto o, s \mapsto S(o-\sigma(now))]$, where
    $S$ is the solution of the continuous with respect to $\sigma(s)$ as defined in the rule.
   Moreover, for any $o \in [\sigma(now), \sigma(now)+d)$,
    $h(o) \models B$, and $\sigma' \models \neg B$ or there exists $\delta>0$ such that
    for any $o \in (\sigma'(now), \sigma'(now)+\delta)$,
    $\sigma'[now \mapsto o, s\mapsto S'(o-\sigma'(now))] \models \neg B$,
    where
    $S'$ is the solution of the continuous with respect to $\sigma'(s)$ as defined in the rule.
    Obviously, $\sigma' \models (\exists s. \Pre) \wedge \close{\neg B}$. According to the definition
    of $Inv$, then for any $o \in [\sigma(now), \sigma(now)+d)$,
    $h(o) \models Inv$, thus
    $\sigma' \models \close{Inv}$ and $h, [\sigma(now),$
    $\sigma'(now)]  \models \dceil{Inv}^<$ hold.
    Plus the fact that $h, [\sigma(now),$
    $\sigma'(now)] \models \dceil{(\exists s. \Pre) \wedge B}^<$, the result is proved.

    \item Case $\bexempt{\evolution{F}{s}{B}}{b}{Q}$: According to the transition system, there are two
    cases for termination, by applying the fourth and the third transition rules for it respectively.
    For the first case, there must exist $d$ such that $\sigma'(now) = \sigma(now) + d$,
    and for any variable $x$ except for $s$ and the ones in $mv(b)$, $\sigma'(x) = \sigma(x)$
    and for any $o\in [\sigma(now), \sigma(now)+d]$, $h(o)(x) = \sigma(x)$. Plus the semantics of
    continuous, we have
    $\sigma' \models \exists mv(b). (\exists s. \Pre) \wedge \close{\neg B} \wedge \close{Inv}$
    and $h, [\sigma(now), $
    $\sigma'(now)]  \models \dceil{\exists mv(b). (\exists s. \Pre )\wedge B \wedge Inv}^<$ proved.
    For the second case, there must exist $d_1$ such that $\sigma''(now) = \sigma(now) + d_1$,
    and for any variable $x$ except for $s$ and the ones in $mv(b)$, $\sigma''(x) = \sigma(x)$
    and for any $o\in [\sigma(now), \sigma(now)+d]$, $h'(o)(x) = \sigma(x)$, and
    $\sigma'' \models (\exists mv(b). (\exists s. \Pre) \wedge \close{Inv}) \wedge\alpha$, and $(Q, \sigma'') \rightarrow (\pstop, \sigma', h'')$, and
    $h = h' {}^\chop h''$. The fact is proved based on the inductive hypothesis on $Q$.

    \item Cases $P\|Q$, $P;Q$ and $\omega \rightarrow P$: According to the transition
    system, for $P\|Q$, suppose $P$ and $Q$ terminate at the same time, then
    there must exist $\sigma_1, h_1$,  and $\sigma_2, h_2$ such that
    $(P, \sigma) \rightarrow (\pstop, \sigma_1, h_1)$, $(Q, \sigma) \rightarrow (\pstop, \sigma_2, h_2)$,
    $\sigma' = \sigma_1 \uplus \sigma_2$ and $h = h_1 \uplus h_2$.
    The fact is proved by induction hypothesis on $P$ and $Q$.  The other cases can be proved
    easily.

    Similarly, the rules for $P;Q$ and $\omega \rightarrow P$ can be proved by induction hypothesis, and we omit
    the details here.

    \item Case $P^*$: According to the transition system,
  we have \[\sigma' = \sigma \quad h = \{\sigma(now) \mapsto \sigma'\}\]
  or there exist an integer $k>0$ such that
    $\sigma_k = \sigma'$, $h = h_1 {}^\chop h_2 {}^\chop \cdots {}^\chop h_k$, and a sequence of
    transitions as follows:
    \[
    \begin{array}{c}
      (P, \sigma) \rightarrow (\pstop, \sigma_1, h_1)\\
      (P, \sigma_1) \rightarrow (\pstop, \sigma_2, h_2)\\
      \cdots\\
      (P, \sigma_{k-1}) \rightarrow (\pstop, \sigma_k, h_k)
    \end{array}\]
    For the first case, the fact holds trivially. For the second case,
    suppose the fact holds when $k<n$ for some $n>0$, next we prove  that
    the fact holds for $k=n$. According to the transition rule,
    we have
    \[
    \begin{array}{l}
     (P, \sigma_{n-1}) \rightarrow (\pstop, \sigma_n, h_n),
      \quad \sigma_{n-1} \models\Pre\\
      \quad h_1 {}^\chop \cdots {}^\chop h_{n-1}, [\sigma(now), \sigma_{n-1}(now)]  \models Inv \vee \ell=0
    \end{array}\]
    By induction hypothesis on $P$,
    $\sigma_n \models\Pre$ and $h_n, $
    $[\sigma_{n-1}(now), \sigma_n(now)]\models Inv$ must hold. Then
    $h_1 {}^\chop \cdots {}^\chop h_n,$
    $ [\sigma(now), \sigma_n(now)] \models (Inv \vee \ell=0)^\chop Inv$, plus $Inv ^\chop Inv \Rightarrow Inv$,
    we have $h_1 {}^\chop \cdots {}^\chop h_n, [\sigma(now), \sigma_n(now)]  \models Inv$ proved.
  \end{itemize}
$\qed$

\subsection{Application: Reachability Analysis}
\label{subsection:reachability}
The inference system can be applied directly for reachability analysis.
Given a labelled process $S$ (a process annotated with integers denoting locations), a precondition $\Pre$ and a location $l$ in $S$, by applying the inference system,
we can deduce a property  $ \psi$ such that if $S$ reaches $l$, $\psi$ must hold at $l$, denoted by
$\typeb{S, l, \Pre}{\psi}$. In another word, If $\typeb{S, l, \Pre}{\psi}$  and $\psi$ is not satisfiable, then $l$
will not be reachable in $S$  with respect to $\Pre$.
We have the following facts based on the structural induction of $S$:
\begin{itemize}
  \item for any process $P$, $\typeb{{}^l P, l, \Pre}{\Pre}$ and $\typeb{P{}^l, l, \Pre}{\Post}$ provided $\inference{\Pre}{P}{\Post, -}$;
  \item $\typeb{\bexempt{\evolution{F}{s}{B}}{{}^lb}{S'}, l, \Pre}{\Pre}$.
  $\typeb{\bexempt{\evolution{F}{s}{B}}{b{}^l}{S'}, l, \Pre}{(\exists mv(b).(\exists s.\Pre) \wedge \close{Inv}) \wedge \alpha}$ (denoted by $\Pre'$), if  $\typeb{b}{\alpha}$ holds.  $\typeb{\bexempt{\evolution{F}{s}{B}}{b}{S'}, l, \Pre}{\psi} $ if $l \in S'$ and  $\typeb{S', l, \Pre'}{\psi}$ hold;
 \item $\typeb{S_1; S_2, l, \Pre}{\psi} $ if $l \in S_1$ and $ \typeb{S_1, l, \Pre}{\psi}$ hold.
 $\typeb{S_1; S_2, l, \Pre}{\psi'} $ if $l \in S_2$,  $\inference{\Pre}{S_1}{\Post, -}$ and $ \typeb{S_2, l, \Post}{\psi'}$ hold;
     \item   $\typeb{ \omega{}^l \rightarrow S', l, \Pre}{\Pre \wedge\omega}$. $\typeb{ \omega\rightarrow S', l, \Pre}{\psi}$ if $l\in S'$ and $\typeb{S', l, \Pre \wedge \omega}{\psi}$;
  \item $\typeb{S'^*, l, \Pre}{\psi}$, if $l\in S'$, $\typeb{S', l, \Pre} {\psi}$ and $\inference{\Pre}{S'}{\Pre, -}$ hold.
\end{itemize}
Obviously, the monotonicity  holds: if $\typeb{S, l, \Pre}{\psi}$ and $\psi \Rightarrow \psi'$, then $\typeb{S, l, \Pre}{\psi'}$.

\example
Consider $P_0$ in Example 2. Given precondition $\Pre$ , we have
$\typeb{P_0, 1, \Pre} {(\exists t. \Pre) \wedge t=0}$,
denoted by $\Pre_1$. Moreover,
$\typeb{P_0, 5, \Pre}{ (\exists mv(b_0). (\exists s, v, t. \Pre_1) \wedge t\leq T) \wedge (u_a=1 \vee w_a=1)
\wedge (u_a=0 \wedge w_a=0)}$, the formula  is un-satisfiable, thus location 5 is not reachable.
Other locations can be considered similarly. $\qed$

\paragraph{\textbf{Implementation}}
We have mechanized the whole framework  in Isabelle/HOL and implemented an interactive theorem prover for
reasoning about hybrid systems modeled using HCSP with binders \footnote{The prover, plus the models and proofs related to the train control example in next section,  can be found at \url{https://github.com/wangslyl/hcspwithbinders}.}.

\section{Train Control Example}
\label{sec:application}
We apply our approach to the train control system  depicted in Fig.~\ref{fig:idea}: firstly, we
construct the formal model for the whole system, especially the train;
secondly,  prove for the train that it is safe against denial-of-service security attack  with respect
to properties (F1) and (F2); finally, explore  the constraints that
relate the constants of different components and learn more precise behavior of
the train.
Assume for the train that
its acceleration ranges over $[-c, c]$ for some $c>0$, and the maximum speed
is  $v_{max}$.

\begin{table}[t]
\small
\centering
\[\begin{array}{lll}
  \textsf{TR} &=& \textsf{MV}(t_1, T_1) \unrhd {}^0 \&_\exists(\textsf{trd}!v\{u_v\}, \textsf{trv}!v\{w_v\}){}^{7}\\
  && \rightarrow  (u_v=1 \wedge w_v=1 \rightarrow (\textsf{MV}(t_2, T_2) \unrhd\&_\exists (\textsf{dr}?x_a\{u_a\}, \textsf{vc}?y_a\{w_a\}) \rightarrow\\
  &&\qquad\ (w_a = 1 \rightarrow  (VA(v, y_a) \rightarrow a: = y_a; \neg VA(v, y_a) \rightarrow \textsf{SC});\\
  &&\qquad\ \ u_a = 1\wedge w_a = 0 \rightarrow (VA(v, x_a) \rightarrow a: = x_a; \neg VA(v, x_a) \rightarrow \textsf{SC});\\
  &&\qquad\ \ u_a = 0 \wedge w_a = 0 \rightarrow  {}^2 \pskip);\   t_2\geq T_2 \rightarrow \textsf{SC};\\
  &&\quad \ u_v=1 \wedge w_v=0  \rightarrow (\textsf{MV}(t_2, T_2) \unrhd\&_\exists (\textsf{dr}?x_a\{u_a\}) \rightarrow\\
  &&\qquad\ (u_a = 1 \rightarrow (VA(v, x_a) \rightarrow a: = x_a;\neg VA(v, x_a) \rightarrow \textsf{SC});  \\
    &&\qquad\quad  u_a = 0 \rightarrow  {}^3\pskip); t_2\geq T_2 \rightarrow \textsf{SC};\\
   &&\quad \ u_v=0 \wedge w_v=1  \rightarrow  (\textsf{MV}(t_2, T_2) \unrhd\&_\exists (\textsf{vc}?y_a\{w_a\}) \rightarrow\\
  &&\qquad \ (w_a = 1 \rightarrow (VA(v, y_a) \rightarrow a: = y_a; \neg VA(v, y_a) \rightarrow \textsf{SC}); \\
    &&\qquad\quad  w_a = 0 \rightarrow {}^4 \pskip); t_2\geq T_2 \rightarrow \textsf{SC};\\
   && \quad\ u_v=0 \wedge w_v=0  \rightarrow  {}^1 \pskip); t_1\geq T_1 \rightarrow \textsf{SC};\\
 \multicolumn{3}{l}{ \textsf{MV (t, T)} =  t:=0; \langle \dot{s} = v, \dot{v} = a, \dot{t} = 1 \& t<T \rangle}\\
  \textsf{SC} &= & a:=-c; \langle \dot{s} = v, \dot{v} = a \&  v>0 \rangle; a:=0
\end{array} \]
\caption{The model of \textbf{train}}
\label{trainmodell}
\end{table}

\begin{table}[t]
\small
\begin{minipage}{0.5\linewidth}
\centering
\[\begin{array}{lll}
\textsf{DR} &= &\pwait\ T_3; {}^5 \&_\exists \textsf{trd}?v_d\{u_v\};  \ {}^{8} u_v = 1\\
&& \rightarrow(v_d \geq (v_{max} - cT_1 -cT_2)\\
&& \qquad  \rightarrow \talloblong_{l\in [-c, 0)} d_a:= l; \\
&&  \  \ v_d < (cT_1 + cT_2) \rightarrow \talloblong_{l\in [0, c]} d_a:= l;\\
&&  \  \ v_d \in [cT_1 + cT_2, v_{max} - cT_1 -cT_2)  \\
&& \qquad \qquad   \rightarrow\talloblong_{l\in [-c, c]} d_a:= l;\\
&&  \ \ \&_\exists (\textsf{dr}!d_a\{u_a\}, \textsf{tick}?o\{u_c\}) \rightarrow\\
&&   \qquad{}^{12}(u_a = 1 \wedge u_c = 1 \rightarrow \pskip;\\
&& \quad  \qquad\ u_a = 1 \wedge u_c = 0 \rightarrow \textsf{tick}?o\{u_c\};\\
&& \quad\qquad\  u_a = 0 \wedge u_c = 1 \rightarrow \pskip;\\
&& \quad\qquad\  u_a = 0 \wedge u_c = 0 \rightarrow \pskip)\\
&& \quad \  \| \textsf{CK}
);\\
&& \ u_v = 0 \rightarrow \pskip\\
\textsf{CK} &=& \pwait\ T_5; \textsf{tick}!\checkmark
\end{array} \]
\end{minipage}
\begin{minipage}{0.5\linewidth}
\centering
\[\begin{array}{lll}
\textsf{VC} &= &\pwait\ T_4; {}^{6} \&_\exists \textsf{trv}?v_r\{w_v\}; {}^{9} w_v = 1 \\
&& \rightarrow(v_r \geq (v_{max} - cT_1 - cT_2)\\
&& \qquad \rightarrow  r_a:= -c; \\
&&\ \ v_r <(cT_1 + cT_2) \rightarrow  r_a:= c;\\
&&  \  \ v_r \in [cT_1 + cT_2, v_{max} - cT_1 -cT_2)  \\
&& \qquad \qquad  \rightarrow \talloblong_{l\in [-c, c]} r_a:= l;\\
&& \ \&_\exists (\textsf{vc}!r_a\{w_a\}, \textsf{tick}?o\{w_c\}) \rightarrow\\
&&  \quad \quad(w_a = 1 \wedge w_c = 1 \rightarrow \pskip;\\
&&  \quad  \quad\ w_a = 1 \wedge w_c = 0 \rightarrow \textsf{tick}?o\{w_c\};\\
&&  \quad\quad\  w_a = 0 \wedge w_c = 1 \rightarrow \pskip;\\
&& \quad\quad\  w_a = 0 \wedge w_c = 0 \rightarrow \pskip)\\
&& \quad \  \| \textsf{CK}
);\\
&&\ w_v = 0 \rightarrow \pskip \\
\\
\end{array} \]
\end{minipage}
\caption{The models of \textbf{driver} and \textbf{VC}}
\label{driverevc}
\end{table}

\paragraph{\textbf{Models}}
The model of the train is given in Table~\ref{trainmodell}. There are two auxiliary processes:
given a clock variable $t$ and  time  $T$,
$\textsf{MV (t, T)}$ defines that the train moves with velocity $v$ and acceleration $a$
for up to $T$ time units;
and $\textsf{SC}$ defines the feedback control of the train when the services from
the driver or the VC fail: it performs
an emergency brake by setting
  $a$ to be $-c$ ,  and
as soon as $v$ is reduced to $0$, resets $a$ to be $0$, thus the train keeps still finally.
The main process \textsf{TR}
models the movement of a train. The train first moves for at most $T_1$ time units,
during which it is always ready to
send $v$ to the driver as well as the VC along \textsf{trd} and \textsf{trv} respectively.
If neither of them responses within $T_1$, indicated by $t_1 \geq T_1$,
the self control is performed.
Otherwise, if at least one communication occurs,
the movement is interrupted and a sequence of
case analysis is followed to execute.

The first case, indicated by $u_v=1$ and $w_v=1$,
represents that the driver as well as the VC succeed to receive $v$.
The train will wait for at most $T_2$ time units
for
the new acceleration from the  driver or the VC along \textsf{dr} and \textsf{vc} respectively, and during the waiting time, it
continues to move with the original acceleration. The new acceleration is expected to satisfy a
 safety condition $VA(v, a)$:
 \vspace{-0.5em}
 \[\begin{array}{l}
 (v > v_{max} - cT_1 - cT_2 \Rightarrow -c \leq a <0) \wedge
  (v < cT_1 + cT_2 \Rightarrow c \geq a \geq 0) \\
 \wedge(cT_1 + cT_2 \leq v \leq v_{max} - cT_1 - cT_2) \Rightarrow (-c \leq a \leq c)
 \end{array}\]
which implies the boundaries for setting $a$ to be positive or negative and is necessary for keeping the velocity always in
$[0, v_{max}]$, otherwise, it will be rejected by the train.
 If both the driver and the VC fail to response within $T_2$, indicated by $t_2 \geq T_2$,
 the self control is performed.
 Otherwise, the following case analysis is taken:
 If the train receives a value (i.e. $y_a$) from  VC,
 indicated by $w_a=1$, then sets $y_a$
 to be the acceleration if it satisfies $VA$,
otherwise, performs self control; if the train receives a value (i.e. $x_a$) from the driver but not from
the VC, updates the acceleration similarly as above; if
the train receives no value from both (in fact never reachable), the skip is performed.


The  other three cases, indicated by $u_v=1 \wedge w_v=0$, $u_v=0 \wedge w_v=1$, and $u_v=0 \wedge w_v=0$,
can be considered similarly.


One possible implementation for driver and VC is given in Table~\ref{driverevc}, in which
process $\pwait \ T_i$ for $i=3, 4$ is an abbreviation for
$t_i :=0;  \langle \dot{t_i} = 1 \& t_i < T_i\rangle$. In process
$\textsf{DR}$, the driver asks the velocity of the train every $T_3$ time units, and as soon as
it receives $v_d$, indicated by $u_v=1$, it computes the new acceleration as follows:
 if $v_d$ is almost reaching $v_{max}$ (by the offset $cT_1 + cT_2$), then chooses a negative
in $[-c, 0)$ randomly; if
$v_d$ is almost reaching 0, then chooses a non-negative in $[0, c]$  randomly;
otherwise, chooses one in $[-c, c]$ randomly.
 The train then sends the value being chosen (i.e. $d_a$) to the train, and if it fails to reach the train
within $T_5$ (i.e. the period of the clock), it will give up. The auxiliary process
\textsf{clock} is introduced to prevent deadlock caused by the situation when the driver succeeds to receive velocity $v_d$ from the train but fails to send acceleration $d_a$ to the train within a reasonable time (i.e. $T_5$ here).
\textsf{VC}  and $\textsf{DR}$ have very similar structure, except that
\textsf{VC} has a different period $T_4$, and it
will choose $-c$ or $ c$ as the acceleration for the first two critical cases mentioned above.

Finally, the train control system can be modeled
as the  parallel composition:
$\textsf{SYS} = \textsf{TR}^* \| \textsf{DR}^* \| \textsf{VC}^* \| \textsf{CK}^*$.
By using $^*$,  each component will be executed repeatedly.

\paragraph{\textbf{Proofs of Train}}
\label{subsec:proof}
First of all,
we  define the precondition of $\textsf{TR} ^*$, denoted by $\Pre_0$, to be
$VA(v, a) \wedge 0 \leq v \leq v_{max} \wedge -c \leq a \leq c$, which indicates that
in the initial state, $v$ and $a$ satisfy the safety condition and are both well-defined.

Secondly, we need to calculate the differential invariants for differential equations occurring in \textsf{TR}.
Consider the equation in $\textsf{MV}(t_1, T_1)$,
the precondition of it with respect to $\Pre_0$, denoted by $\Pre_1$, can be simply calculated, which is $\Pre_0 \wedge t_1=0$, then
by applying the method proposed in~\cite{LZZ11}:
\[\left(
\begin{array}{l}
  \big(0 \leq t_1 \leq T_1\big) \bigwedge \\
   \big(a < 0  \Rightarrow (v \geq cT_2 + (at_1 + cT_1)) \wedge
    (v \leq v_{max}) \big)\\
   \bigwedge \big(a \geq 0  \Rightarrow (v \leq v_{max} -cT_2 + (at_1 - cT_1)) 
  \wedge (v \geq 0) \big)
\end{array}\right)\]
denoted by $Inv_1$, constitutes a differential invariant of the continuous with respect to $\Pre_1$.
It is a conjunction of three parts, indicating that:
(1) $t_1$ is always in the range $[0, T_1]$;  (2) if $a$ is negative, $v$
must be greater or equal than $cT_2$ plus  a positive value (i.e. $at_1 + cT_1$), and meanwhile $v \leq v_{max}$;
and (3) if $a$ is positive,
$v$  must be less or equal than $v_{max} -cT_2$ plus a negative value (i.e. $at_1 - cT_1$), and meanwhile $v \geq 0$.
This invariant is strong enough for guaranteeing  $cT_2 \leq v \leq v_{max}-cT_2$ after the continuous escapes no matter what $a$
is in $[-c, c]$.
Similarly, we can calculate the invariant of the continuous occurring in $\textsf{MV}(t_2, T_2)$, which is
\[\left(
\begin{array}{l}
  \big(0 \leq t_2 \leq T_2\big) \bigwedge \\
   \big(a < 0  \Rightarrow (v \geq 0 + (at_2 + cT_2)) \wedge
    (v \leq v_{max}) \big)\\
   \bigwedge \big(a \geq 0  \Rightarrow (v \leq v_{max} + (at_2 - cT_2))
  \wedge (v \geq 0) \big)
\end{array}\right)\]
denoted by $Inv_2$. This invariant is strong enough for guaranteeing $0 \leq v \leq v_{max}$ after the continuous escapes.
Finally, the invariant of the differential equation of $\textsf{SC}$ is $0 \leq v \leq v_{max}$, and we denote it by $Inv_3$.

Next, to prove (F1) and (F2), we can prove the following facts instead:
\begin{itemize}
\item Locations 1, 2, 3, 4 are not reachable for $\textsf{TR}^*$;
\item Throughout the execution of $\textsf{TR}^*$,
the invariant $ 0 \leq v \leq v_{max}$ always holds.
\end{itemize}
First we consider one loop of execution $\textsf{TR}$.
For location 1, we can deduce that \footnote{For simplicity, we use the boldface of an acknowledgment  variable to represent the corresponding
  formula, e.g., $\textbf{u}_\textbf{v}$ for $u_v=1$.}
$\typeb{\textsf{TR}, 1, \Pre_0} {(\textbf{u}_\textbf{v} \vee \textbf{w}_\textbf{v}) \wedge (\neg \textbf{u}_\textbf{v} \wedge \neg \textbf{w}_\textbf{v})}$, which
is not satisfiable, thus location 1 is never  reachable.
Similarly, we can deduce that locations 2, 3, 4 are not reachable as well.  %
%
Furthermore, by applying the inference system, we can deduce the  specification
$\inference{\Pre_0}{\textsf{TR}}{\Pre_0, \lceil 0 \leq v \leq v_{max} \rceil^<} $.
After one loop  of execution of the train, $\Pre_0$ still holds at termination. Thus,
all the above reachability  results obtained for $\textsf{TR}$ still hold for $\textsf{TR}^*$,  whose execution
is equivalent to some  finite number of executions of  $\textsf{TR}$.
Finally, plus that $\lceil 0 \leq v \leq v_{max} \rceil^<$ is idempotent over chop, we
can deduce
$\inference{\Pre_0}{\textsf{TR}^*}{\Pre_0, \lceil 0 \leq v \leq v_{max} \rceil^<}$, denoted by (\textbf{TrainSpec}),
which
implies that $0 \leq v \leq v_{max}$ is an invariant for the train.

By applying  our interactive theorem prover, the fact (\textbf{TrainSpec}) is proved as a theorem, and
the above reachability results can be implied from the lemmas proved for corresponding processes, according to
the method introduced in Section~\ref{subsection:reachability}.

We can see that, most of the proofs need to be performed in an interactive way,
mainly because of the following reasons: firstly, we need to provide the differential invariants by ourselves during proof
of continuous evolution; and
secondly, we need to conduct the proof of DC formulas by telling
which axiom or inference rule of DC should be applied.  For the first problem, we will consider
the integration of the prover to a differential invariant generator that can be implemented based on the method
proposed in~\cite{LZZ11}. For the second, we will consider  the decidability
of DC and design algorithms for solving the decidable subsets, or as an alternative approach,
consider translating DC formulas into HOL formulas in a semantic way and
applying the existing automatic solvers for HOL instead. Both of these will be our future work.

\paragraph{\textbf{Constraints of Constants}}

We can further analyze the behavior of the whole system $\textsf{SYS}$. By defining the constraints relating
 different constants, the behavior of communications between
the three components can be determined.
Consider the first loop of execution of each component, based on reachability analysis, we have the following facts:
for locations 0, 5, 6,  $t_1=0$, $t_3 = T_3$ and $t_4 = T_4$ hold respectively, and
for locations 7, 8, 9, $t_1 \leq T_1$, $t_3 \geq T_3$ and $t_4 \geq T_4$ hold respectively.
The synchronization points have four possibilities: $(7, 8)$,  $(7, 9)$,  $(7, 8, 9)$, or none. For the first case,
i.e. the train succeeds to communicate with the driver but not with the VC,
there must be $t_1 = t_3 < t_4$, and if $T_3 < T_4$ and $T_3 \leq T_1$ hold, this case will occur.
The second one is exactly the contrary case.
For the third case, there must be $t_1 = t_3 = t_4$, and if
$T_3 = T_4 \leq T_1$ holds, this case will occur. Finally, if both
$T_3 > T_1$ and $T_4 > T_1$ hold, the last case occurs, i.e., locations 7, 8 and 9 are not reachable, and thus
the train fails to communicate with both the driver and the VC.
Following this approach,  more precise behavior of the communications of the train
can be obtained.

\section{Conclusion and Future Work}
\label{sec:conclusion}

This paper proposes a formal modeling  language, that is a combination of hybrid CSP and binders from quality calculus,
 for expressing denial-of-service due to unreliable communications in hybrid systems.
With the linguistic support, it is able to build a safe hybrid system that behaves in a reasonable manner
in the presence of denial-of-service security attack. 
The idea is that, when the service from the controllers  fails, the physical
system itself needs to provide feedback control, in order to meet the safety requirements.
The paper also develops an inference system for reasoning about  such systems,
with no dependence on the behavior of the environment, and furthermore  implements an interactive theorem prover.
We illustrate our approach by considering an example taken from train control system.

The investigation of our approach to more
complex hybrid systems is one of our future work. Meanwhile, for facilitating practical applications, we will consider
to achieve more support of automated reasoning in the theorem prover.

%
\bibliographystyle{plain}
\bibliography{reference}

\end{document}

%% file: pic1.tex

\def\pnode[#1,#2]#3{
\node[pshape,
 text width=1.2cm,
 anchor=center] at #2 (#1)
{#3};
}

\begin{picture}(158,77)(0,0)

\begin{tikzpicture}[->,>=stealth']

\tikzstyle{every node}=[font=\small]

\pnode[Train,(0,0)] {Train};
\pnode[Driver,(-1.5,1.5)] {Driver};
\pnode[VC,(1.5,1.5)] {VC};

\path
(-0.51,0.31)			edge  node[bend left,left]{trd} (-1.35,1.15)
(Driver)		    	edge  node[anchor=south,right]{dr} (Train)
(VC)			edge  node[anchor=south,left]{vc} (Train)
(0.51,0.31)			edge  node[anchor=south,right]{trv} (1.35,1.15);

\end{tikzpicture}

\end{picture}